# From Fault Creep to slow and fast Earthquakes in Carbonates

**François X. Passelègue[1,2], Jérôme Aubry[3], Aurélien Nicolas[3], Michele Fondriest[2], Damien Deldicque[3], Alexandre Schubnel[3] and Giulio Di Toro[2,4]**

[1] *Laboratoire Expérimental de Mécanique des Roches, École Polytechnique Fédérale de Lausanne, Lausanne, Switzerland*

[2] *SEES, The university of Manchester, Manchester England*

[3] *Laboratoire de Géologie de l'École Normale Supérieure, Paris, France*

[4] *Dipartimento di Geoscienze, Università degli Studi di Padova, Padua, Italy*

**ABSTRACT**

**A major part of the seismicity striking the Mediterranean area and other regions worldwide is hosted in carbonate rocks. Recent examples are the destructive earthquakes of L'Aquila $M_w$ 6.1 2009 and Norcia $M_w$ 6.5 2016 in Central Italy. Surprisingly, within this region, fast (≈3km/s) and destructive seismic ruptures coexist with slow (≤10 m/s) and non-destructive rupture phenomena. Despite of its relevance for seismic hazard studies, the transitions from fault creep to slow and fast seismic rupture propagation are still poorly constrained by seismological and laboratory observations. Here, we reproduced in the laboratory the complete spectrum of natural faulting on samples of dolostones representative of the seismogenic layer in the region. The transitions from fault creep to slow ruptures and from slow to fast ruptures, are obtained by increasing both confining pressure (P) and temperature (T) up to conditions encountered at 3-5 km depth (i.e., P = 100 MPa and T = 100 ºC), which corresponds to the hypocentral location of slow**

**earthquake swarms and the onset of regular seismicity in Central Italy. The transition from slow to fast rupture is explained by the increase of the ambient temperature, which enhances the elastic loading stiffness of the fault and consequently the slip velocity during the nucleation stage, allowing flash weakening. The activation of such weakening induces the propagation of fast ruptures radiating intense high frequency seismic waves.**

## INTRODUCTION

In the Earth's upper crust, faults release elastic strain energy stored in the wall rocks via different modes of slip. Depending on the velocity of the rupture front ($V_r$), faults may creep, or generate slow ($V_r \leq 10$ m/s) (Ide et al., 2007), regular ($V_r \approx 3000$ m/s also called fast ruptures) and supershear ($V_r \geq 4200$ m/s) earthquakes (Kanamori and Brodsky, 2004; Bouchon and Vallée, 2003; Passelègue et al., 2016). The Mediterranean area and several other regions worldwide are affected by moderate to large in magnitude earthquakes (Chiaraluce et al., 2012, Vaoloroso et al., 2013) nucleating and propagating within a (4-8 km) carbonate sequences (i.e., limestones and dolostones). This is the case of the Northern and Central Apennines of Italy which were recently struck by destructive seismic sequences largely hosted within dolomitic rocks (Chiaraluce et al., 2012, Vaoloroso et al., 2013) (Figs. 1a and 1b). These sequences were characterized by complex mainshocks-aftershocks spatio-temporal distributions with (i) most of the seismicity compartmentalized between 10 and ≈3 km depth and (ii) a sharp upper seismicity cut-off at ≈3 km depth (Chiaraluce et al., 2012, Vaoloroso et al., 2013) (Fig. 1b). Remarkably, in this region destructive fast seismic ruptures coexist with slow (≤10 m/s) and non-destructive rupture phenomena (Crescentini et al., 1999; Amoruso et al., 2002) (Figs. 1a and 1b). From a rock mechanics point of view, the coexistence of these different modes of slip remains unclear because carbonates can accommodate deformation by ductile crystal-plastic processes, such as aseismic mechanical twinning and dislocation glide, even at room temperature (De Bresser and Spiers, 1997). The ability of calcite crystals to deform plastically at low pressure and temperature can explain the lack of acoustic emission activity (micro-seismicity) during failure of carbonates under shallow depth conditions (Schubnel et al., 2006; Nicolas et al., 2017). Indeed recent experimental studies focused on the frictional behavior of calcite- and dolomite-rich fault rocks sheared from (i) sub-seismic slip rates ($V_s$<0.1 mm/s) to determine the rate and state friction parameters at ambient (Scuderi et al., 2013) and crustal temperatures (Verberne et al., 2015), to (ii) seismic slip rates (0.01< $V_s$ <10 m/s) to understand the weakening processes of carbonates under different environmental conditions (Fondriest et al., 2013; De Paola et al., 2015). Here we report the results of triaxial experiments performed on saw cut samples cored in dolostone blocks of the Mendola Formation (Upper Triassic in age), a dolomitized platform carbonate with crystal size in the range 20 to 300 µm and larger crystals filling diagenetic pores (Fondriest et al., 2015). The experiments reproduced the ambient conditions (temperature and pressure) typical of the Earth's crust where all these different slip modes occur (Figure 1).

**EXPERIMENTAL METHOD**

The rock cylinders were saw-cut to create an experimental fault at an angle of 30º with respect to $\sigma_1$ (principal stresses are denoted $\sigma_1$>$\sigma_2$≥$\sigma_3$). All experiments were conducted on fault surface presenting the same initial geometry and roughness (grit number #240, 50±10 µm roughness (both spacing and height of asperities)), imposing constant axial strain-rate ($10^{-6}$ or $10^{-5}$ s$^{-1}$, corresponding to slip velocities of 0.1 and 1 micron/s) under 30, 60 and 90 MPa confining pressure and at temperatures of 25, 65 and 100 ºC. Slow rupture phenomena were recorded using four strain gauges equally spaced 1.5 cm appart and recording preferentially shear strain were glued at 4 mm from the fault plane along the fault strike. Fast motions were recorded using 15 piezoelectric transducers sensitive to particles acceleration and a 350 $\Omega$ full bridge strain gauge sensitive to differential stress located close to the fault plane (Passelègue et al., 2016).

**EXPERIMENTAL RESULTS**

At 25 °C, slip initiates when the shear stress reaches the peak strength of the fault, corresponding to a static friction $f_s = \tau_0/\sigma_n^0 \approx 0.4$ (where $\tau_0$ and $\sigma_n^0$ are the shear stress and the normal stress at the onset of instability, respectively). At this temperature, the fault exhibits regular frictional behavior (dependence of confining pressure on peak shear stress) and strain energy accumulated during loading is released by stable slip (Fig. 1c). Increasing the ambient temperature to 65 °C preserves the frictional behavior of the fault, but leads to a transition from stable slip to stick-slip motions (Fig. 1d). In this case, while stress-strain curves suggest regular stick-slip motions, slow ruptures are observed and no high frequency radiations are recorded on piezoelectric transducers. At 100 °C, the fault exhibits a different mechanical behavior. Fast ruptures are observed at both 60 and 90 MPa confining pressure, while only slow ruptures are observed at 30 MPa confining pressure. Fast ruptures induce strong high frequency motions, recorded on both high frequency acoustic and dynamic strain monitoring systems. Note that coexisting slow ruptures, which do not produce high frequency motions, are also observed in the first stage of the experiments conducted at 90 MPa confining pressure. For each event, the peak shear stress at the onset of slip increases with normal stress, which shows that independently of the temperature and pressure conditions, regular frictional behavior leading to fault reactivation is observed (Fig. 2a).

At low confining pressure (30 MPa), the increase of the peak friction with ambient temperature leads to a transition from stable slip to slow rupture. At 60 and 90 MPa confining pressure, a similar trend is observed: an increase in temperature leads to an increase of the peak friction coefficient and to a transition from stable slip to slow rupture, but to fast rupture at T = 100 ºC (Fig. 2a). Moreover, the amount of stress released via the different modes of slip observed depends on the peak shear stress reached during the loading. The larger the peak shear stress, the larger the stress drop (Fig. 2b). However, frictional drop remains small during stable slip and slow rupture ($\Delta f \approx 0.05$). For similar values of initial shear stress, fast rupture releases a larger amount of shear stress (Fig. 2 b), i.e. larger frictional drop, which ranges from 0.07 to 0.22 (Fig. 2b). This behavior is highlighted by comparing the static stress drop of each events to the related amount of slip. Each mode of slip (i.e., stable, slow and fast rupture) presents a different linear relation between the static stress drop and the fault slip (Fig. 2c). For the same value of slip, the resulting stress drop is larger during fast ruptures than during slow ruptures,. Note that slow ruptures observed at both 65 and 100 ºC follow the same trend, suggesting similar mechanisms. These results suggest that fast ruptures are more dispersive or require more energy than slow ruptures.

**NUCLEATION OF SLOW AND FAST RUPTURE**

Using the travel times of the rupture front recorded by the array of strain gauges located along fault strike, estimates for $V_r$ during slow rupture propagation range from 0.1 to 20 m/s (Fig. 3a). These values are in agreement with rupture velocities of natural slow earthquakes (Ide et al., 2007), suggesting that our experimental slow ruptures are similar to those observed in nature. An increase of $V_r$ during rupture propagation is observed along fault strike (Fig. 3a). In addition, increasing the initial shear stress (i.e. the confining pressure) lead to larger rupture velocities at the onset of the frictional instability. To further analyze the influence of the background shear stress on slow rupture propagation, we computed the evolution of stress with slip (weakening stiffness $K=\Delta\tau/\Delta u$) during each event induced during the experiments conducted at 65 ºC under 30, 60 and 90 MPa confining pressure, respectively. The weakening processes of slow rupture is purely slip-weakening (Fig. 3b), confirming recent expectations (Ikari et al., 2013). In addition, the increase of the peak shear stress along fault strike leads to a larger fraction of the shear stress released during the slip events (Fig. 3b). Assuming a typical scheme of earthquake energy budget (Kanamori and Brodsky., 2004), pure slip weakening behavior is expected to drastically limit the radiated energy during rupture propagation, explaining the smaller stress drop for a given amount of slip compared to fast rupture phenomena.

Fast earthquakes radiating high frequency waves present a complex nucleation behavior. At the onset of slip, dynamic strain recording shows that fault slip accelerates and radiates low amplitude and relatively low frequency (20 kHz) acoustic waves (Fig. 3c). At a critical point (red dashed line in Fig. 3c), the shear stress drops abruptly within 20 μs and high frequency waves are radiated during fast rupture propagation. Using piezoelectric transducers as seismic rupture chronometers (Passelègue et al., 2016), we estimate rupture velocities ranging from 1500 to 5200 m/s. Our estimations for ruptures speed are compatible with previous studies (Passelègue et al., 2016) and with the rupture speed of regular earthquakes (Ide et al., 2007). The amplitude and the frequency of the acoustic motions increase with the stress release rate (Figs. 3c and 3d).

**POST MORTEM MICROSTRUCTURES**

Fault surfaces recovered from experiments where stable slip and slow ruptures occurred have highly light reflective patches visible at the naked eye. These patches are very extended and elongated along the slip direction (i.e., parallel to the slicken-lines), and are similar to the mirror-like slip surfaces previously observed in nature or after friction experiments conducted from sub-seismic to seismic slip rates on carbonate rock gouges (Fondriest et al., 2013; Verberne et al., 2015). Scanning electron microscope (SEM) images reveal the extremely smooth topography of mirror surfaces, composed of tightly packed to welded sub-rounded nanograins with negligible porosity (Fig. 4a). Instead, fault surfaces which experienced fast ruptures are dramatically different. First, even though sporadic highly smooth patches with preserved slicken-lines are present, the fault surface is on average much rougher than after stable slip and slow rupture. Second, the fault surface is pervasively covered by a foam-like material embedding small well-rounded nanograins with an average size of 150 nm (Fig. 4b). The foam-like material locally includes ultra-thin (>5 nm in thickness) filaments and patches connecting and wrapping the nanograins (Fig. 4b), recalling frictional melting textures found in silicate-bearing rocks sheared under similar deformation conditions (Passelègue et al., 2016).

**INTERPRETATION AND DISCUSSION**

Our experiments reproduce the complete spectrum of natural faulting: (i) stable slip at room temperature, (ii) slow ruptures at 65 °C and (iii) coexisting slow and fast ruptures at 100 °C. In contrast to previous studies (Leeman et al., 2016), transitions between these different slip modes depend on ambient fault temperature rather than on confining pressure. This finding could be counter-intuitive since the increase of ambient pressure and temperature enhances micro-plasticity in carbonates that should release part of the stored elastic strain energy (Rutter, 1972;

Nicolas et al., 2017). In our experiments, the shear stress increases with normal stress (Fig. 2a), and during dedicated creep experiments, the fault does not exhibit any creep behavior even at 95 % of the peak frictional strength (at 100 ºC). As a consequence, the slip instability is probably driven by frictional rather than crystal-plastic processes. The transition from stable sliding to slow to fast rupture is observed with increasing macroscopic peak friction coefficient (Figs. 2a and 4c), as previously observed (Ben-David et al., 2010). However, large values of peak friction alone are not sufficient to induce fast rupture propagation at low confining pressure so that the weakening rate (as a function of slip) must also play a role in the slip dynamics.

The weakening rate increases with the peak slip rate reached during rupture propagation (Fig. 4d). As expected theoretically, ruptures switch from slow to fast when the fault weakening rate becomes larger than the stiffness of our experimental apparatus (Leeman et al., 2016), i.e. when the nucleation length becomes smaller than the experimental fault (Fig. 4d). Our results also demonstrate that the weakening rate is a logarithmic function of the slip rate, at least for slow and stable slips, as predicted by rate and state friction law (Fig. 4d). For weakening stiffness above that of the apparatus ($K>K_c$), the slip rate seems to saturate at 1 m/s, within the limit of our resolution ($V_s$ is calculated for fast ruptures assuming that the measured slip occurs within the weakening stage (Passelègue et al., 2016)) and the logarithmic relationship between $V_s$ and $K$ breaks down (Fig. 4d), as expected once fault lubrication occurs at seismic slip rates (Goldsby and Tullis, 2011). Indeed, sliding velocities above $V_w$ (≥0.1 m/s) suggest the activation of flash heating during fast events (Goldsby et al., 2011). Flash heating on asperities explains both (i) the reduction of the nucleation length to values smaller than the experimental fault (Fig. 4d) and (ii) the strong enhancement of the weakening rate, which initiates the radiation of high frequency seismic waves (Figs. 3c and 3d). Two mechanisms can potentially operate in the case of carbonate rocks: (i) grain boundary sliding aided by diffusion creep possibly associated to decarbonation (Green et al., 2015; De Paola et al., 2015) or (ii) frictional melting (or a combination of both). Fault surfaces that recorded fast ruptures showed oxides wrapped by a foam-like material that resembles a solidified melt. In carbonate minerals (both calcite and dolomite), the melting point decreases dramatically in the presence of $CO_2$ (Brooker, 1998). Our microstructural observations thus suggest decarbonation enhanced melting along the fault surface (Fig. 4d).

Our experimental approach succeeded in explaining the onset of fault creep and the transition from slow (aseismic) ruptures (Crescentini et al., 1999; Amoruso et al., 2002) to regular seismicity at P-T conditions encountered at 3km depth, as was observed in the Central Apennines based on seismological and geodetic investigations (Chiaraluce et al., 2012). Since all the experiments were performed in dry conditions, it seems that this first-order similarity between experimental and natural fault slip modes was mainly controlled by the variation in confining pressure and temperature. In contrast with silicate-bearing rocks, dolomite and calcite are all expected to behave in a semi-brittle ductile manner (Wong and Baud, 2012; Nicolas et al., 2017) at moderate pressure and temperature conditions. Within the P-T conditions investigated, dolomitic fault interfaces are frictional. Surprisingly, the increase of the ambient temperature with depth promotes velocity weakening behavior in carbonates and the nucleation of fast ruptures. In the temperature range investigated, such behavior has already been observed in granite (Blanpied et al., 1995) and calcite (Verberne et al., 2015). This difference between the rheology of a bulk medium (ductile) and that of an interface (brittle) highlights the importance of local strain rates at asperities, which may control both fault and crustal rheology, as well as potentially trigger earthquakes at or below the brittle ductile transition.

**FIGURE CAPTIONS**

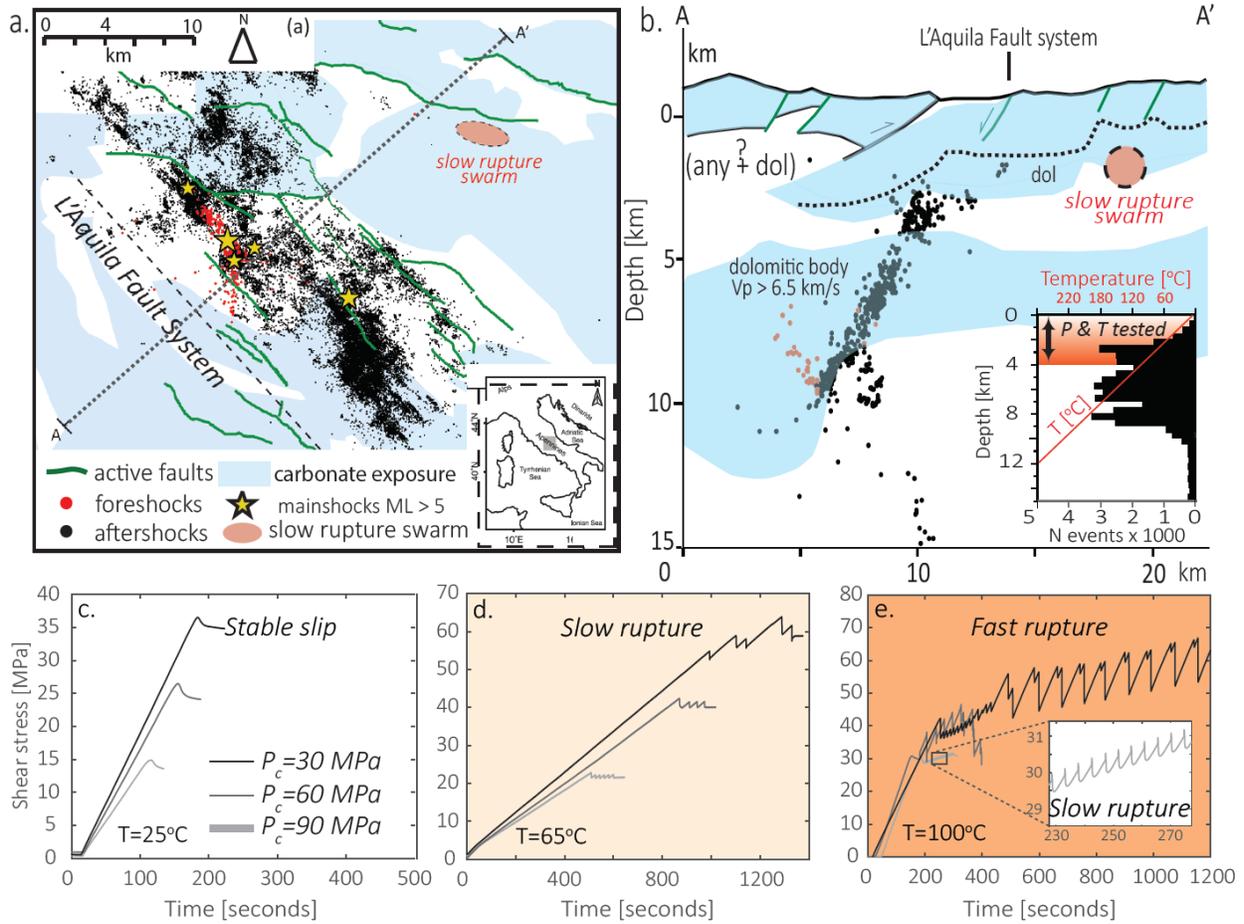

Figure 1. a. Geographical distribution of the seismicity of the L'Aquila 2009 sequence (Valoroso et al., 2013). The surface exposure of carbonate rocks is from published geological maps (Ghisetti and Vezzani, 1998). The approximate location of the slow rupture swarm is derived from geodetic studies (Crescentini et al., 1999; Amoruso et al., 2002). b. Cross section of the L'Aquila fault system (A-A' in Fig. a.) (Ghisetti and Vezzani, 1998; Valoroso et al., 2013). The dolostone body is interpreted from seismic tomography (Di Stefano et al., 2011) and aeromagnetic anomaly measurements (Speranza and Minelli, 2014). The insight in Figure b. displays the distribution with depth of the seismicity associated to the L'Aquila 2009 sequence

(Valoroso et al., 2013). The red line represents the evolution of the temperature within the crust assuming a geothermal gradient of 25 °C/km. The background orange colored area highlights the temperature and pressure conditions tested in the experiments of this study. (c), (d) and (e) show the results of the experiments conducted at increasing confining pressures (30, 60 and 90 MPa) and temperatures (25, 65 and 100 °C).

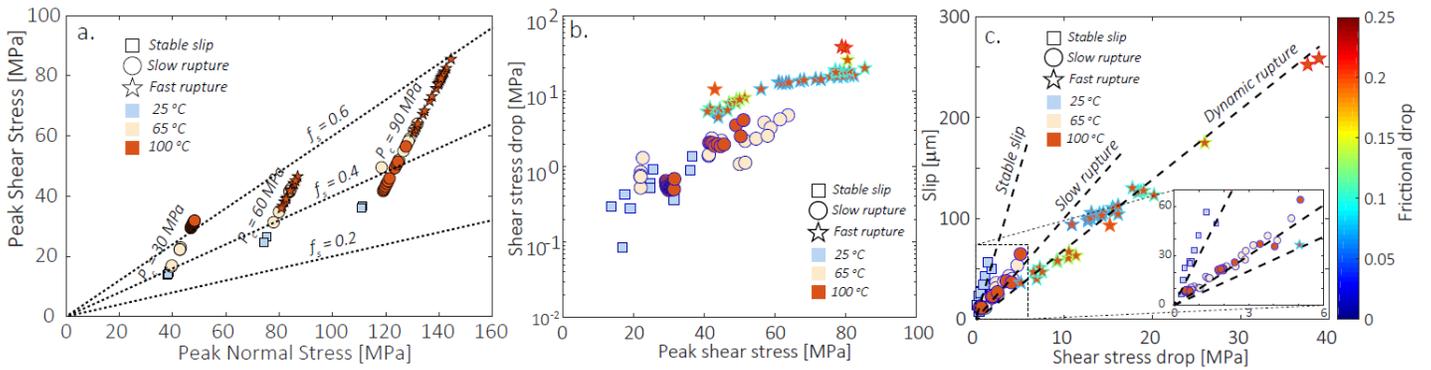

Figure 2. (a) State of stress (shear stress and normal stress) at the onset of slip of each event. Black dashed lines correspond to different values of friction coefficient. (b) Static shear stress drop as a function of the initial peak shear stress. (c) Scaling law of the seismic slip versus the static stress drop for each mode of slip. Black dashed lines correspond to the best linear fit of each mode of slip. The color bar refers to the rim of each symbol in Figs (b) and (c) and indicates the frictional drop during each slip event.

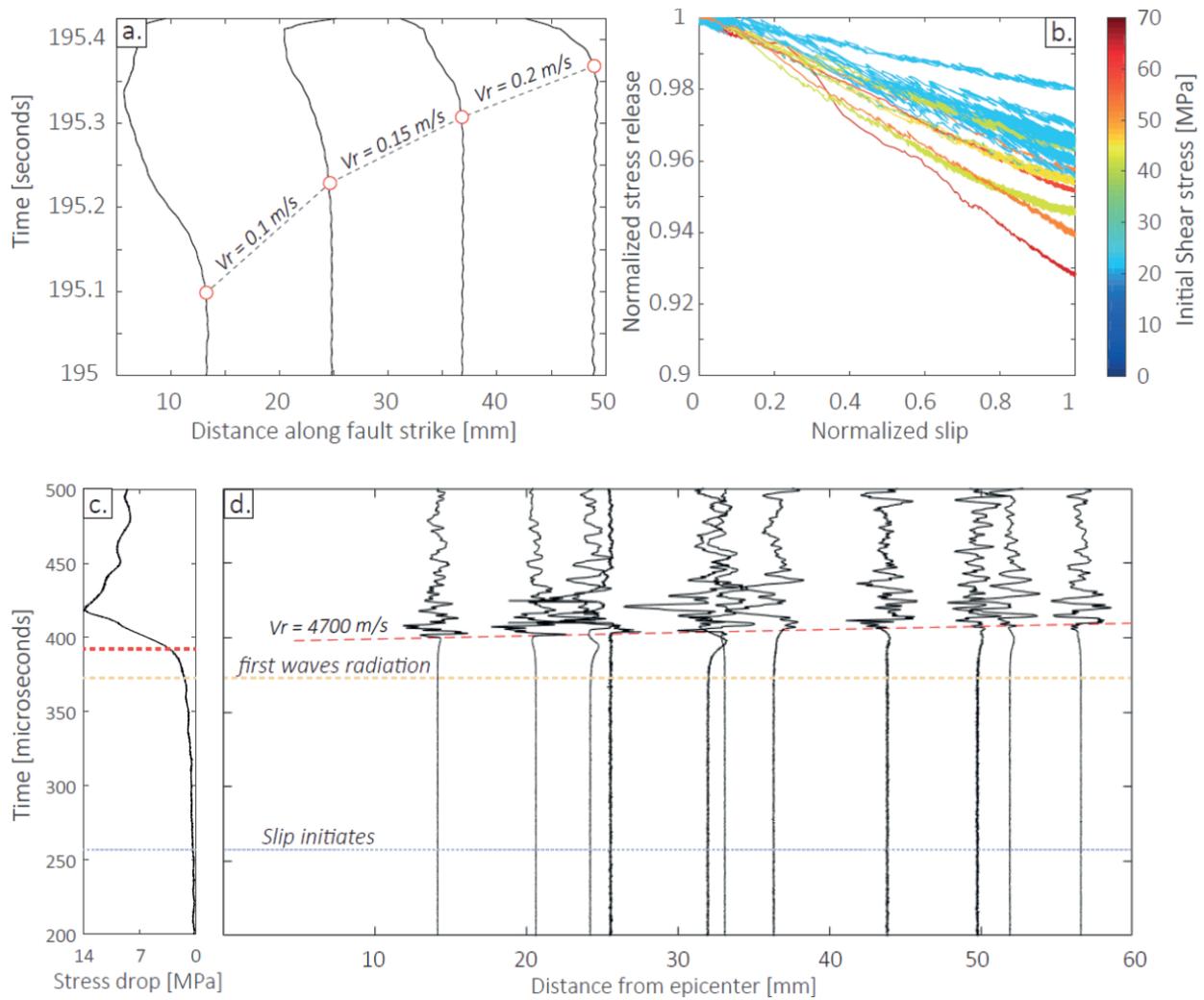

Figure 3. (a) Evolution of the strain on each strain gauge (located along fault strike) as a function of time. The initiation of the strain drop (red circles) on each strain gauges is used to track the rupture front velocity between each strain gauge. (b) Evolution of the shear stress versus slip during all slow rupture events. The slip is normalized by the final slip of each event and the shear stress by the peak shear stress at the onset of slip (presented by the color bar). (c) Evolution of stress during the nucleation and the propagation of a fast rupture. (d) Evolution of acoustic signals during the nucleation and the propagation of a fast rupture as a function of the distance from the initiation of the high frequency motions. The alignment of the high frequency radiation front allows us to estimate the rupture velocity during a slip event.

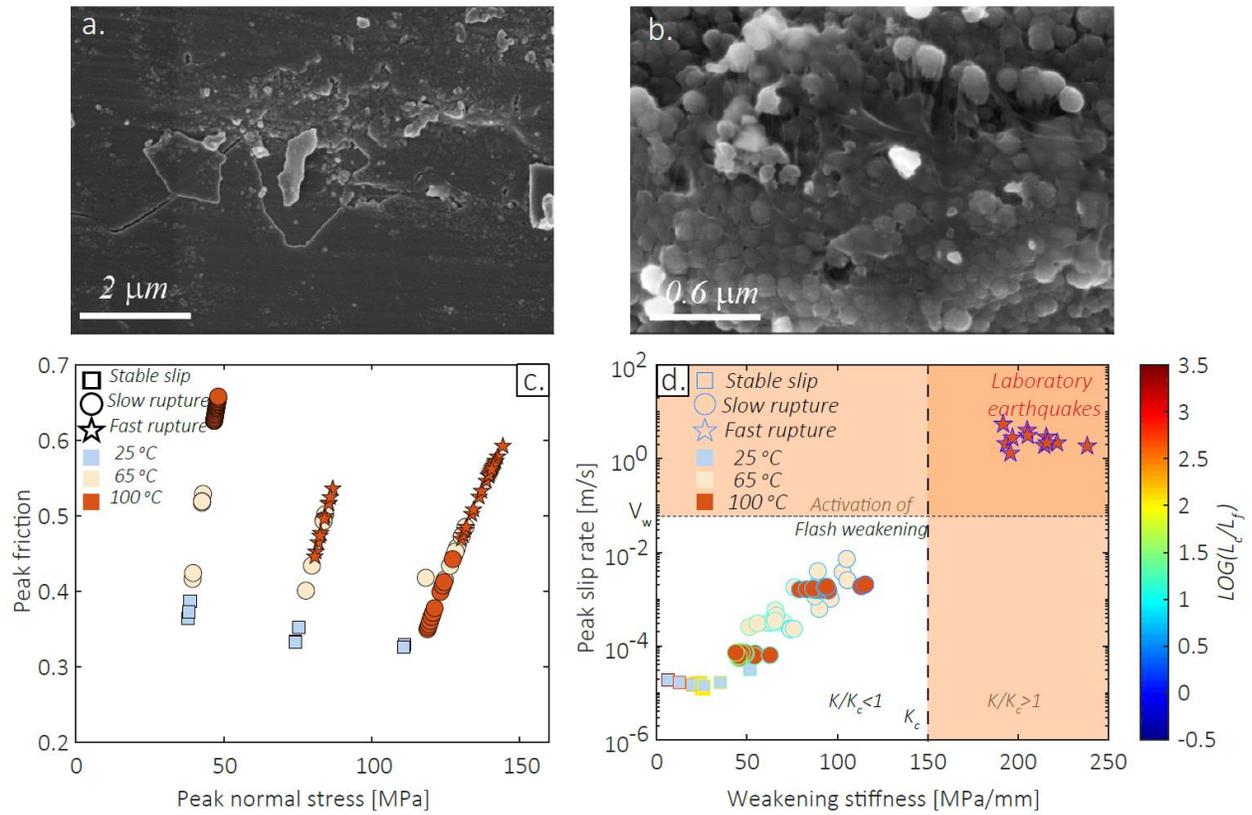

Figure 4. (a) and (b) are SEM images of the fault surfaces which were associated with slow and fast propagating ruptures, respectively. (c) Influence of normal stress and ambient temperature on the peak friction coefficient at the onset of slip and consequences on the rupture mode. (d) Peak slip rate versus weakening stiffness during each event. $K_c$ defines the stiffness of the experimental apparatus. $V_w$ defines the critical velocity for the triggering of flash heating and weakening phenomena at 100 °C ambient temperature, assuming the relation $V_w=(\rho \times C_p)^2(T_w-T_0)^2 \pi \times \kappa/(D(\tau_c)^2)$ where $\rho$ is the rock density (2650 kg/m$^3$), $C_p$ is the heat capacity (900 Jkg$^{-1}$K$^{-1}$), $\kappa$ is the thermal diffusivity (1.25.10$^{-6}$ m$^2$s$^{-1}$), $T_w$ is the decarbonation temperature (600 °C), $T_0$ is the ambient temperature, $D$ is the asperity size (50 and 1 microns) and $\tau_c$ is the contact hardness of dolomite (3 GPa) (Goldsby et al., 2011). The color bar corresponds to the nucleation length of each event normalized by the fault length ($L_f$). Nucleation length was calculated following

$L_c=(2\beta \times \mu \times G)/((\tau 0\times(1-f_d/f_s)^2)$ (Campillo et al., 1997), where $\beta$ is a non-dimensional shape factor coefficient ($\approx$1.158), $\mu$ is the shear modulus of the dolomite estimated using strain measurements, $G$ is the effective fracture energy $G=u\times\sigma_n\times(f_s-f_d)/2$.